Dual-sided transparent display

Suman Halder[1], Yunho Shin[2], Yidan Peng[3], Long Wang[3], Liye Duan[3], Paul Schmalenberg[4], Guangkui Qin[3], Yuxi Gao[3], Ercan M. Dede[4], Deng-Ke Yang[1,2], Sean P. Rodrigues[4,*]

1. Department of Physics, Kent State University, Kent, OH 44242
2. Advanced Materials and Liquid Crystal Institute, Kent State University, Kent, OH 442422
3. Central Research Institute, BOE Technology Group Co., Ltd., Beijing, China
4. Electronics Research Department, Toyota Research Institute of North America, Ann Arbor, MI 48105

In the past decade, display technology has been reimagined to meet the needs of the virtual world. By mapping information onto a scene through a transparent display, users can simultaneously visualize both the real world and layers of virtual elements. However, advances in augmented reality (AR) technology have primarily focused on wearable gear or personal devices. Here we present a single display capable of delivering visual information to observers positioned on either side of the transparent device. This dual-sided display system employs a polymer stabilized liquid crystal waveguide technology to achieve a transparency window of 65% while offering active-matrix control. An early-stage prototype exhibits full-color information via time-sequential processing of a red-green-blue (RGB) light-emitting diode (LED) strip. The dual-sided display provides a perspective on transparent mediums as display devices for human-centric and service-related experiences that can support both enhanced bi-directional user interactions and new media platforms.

## 1. INTRODUCTION

Display systems are undergoing rapid transformations to meet the growing demands of augmented reality (AR). Typically, AR devices are commercialized as wearable technology and offer personalized information to augment the user's view.[1–5] In a similar fashion, though not commonly considered AR tools, transparent display panels offer a window to visualize a scene while also overlapping information on the display[6–8]. To realize the full potential of transparent displays as augmented reality devices, a series of challenges remain. The first nontrivial problem is the tradeoff that exists between brightness and transparency. Current commercial transparent displays only provide transparency ranging from 33-45% for organic light emitting diode (OLED) technologies such as LG's 55EW5PG-S at 38% at 400 nit and OLED Space at 45% and 200 nit,[9–12] 10-20% for polarized liquid crystal display (LCD) panels[13,14], 70-90% for photoluminescent displays[15], and 85% light plate/waveguide devices.[16] Commercial OLED and LCD technologies demonstrate brighter screens, while light plate devices have yet to maximize their scattering efficiencies[17,18] and photoluminescent screens lack multi-color integration on a pixel level. In addition to these specifications, there remains a secondary challenge with transparent displays, known as "bleed-through," that prevents broader market realization.[19,20] Specifically, for a display with a single image portrayed on the front, there exists on the rear side of the display, a mirrored image of the front side that bleeds through. For this reason, the application of a single-image transparent display is limited to areas where only the front side can typically be observed, for instance, in a residential or commercial setting like a shopping window that is furnished with a backdrop. Furthermore, the absence of user privacy poses a significant hindrance, preventing the commercialization of bleed-through transparent displays in both corporate settings and personal smartphone use.[21–25] In order to expand the application of these displays to environments where people may be present on both sides, i.e., most vehicle and building windows, this bleed-through issue must be addressed to realize effective transparent display systems. While transparent, AR displays may be functional in one direction, their purpose and value depreciate when the device cannot be made dual-sided. While this paper does not delve into a comprehensive review of transparent display technologies, readers can discover valuable insights on transparent LCDs, transparent OLEDs, projection technologies, and noteworthy highlights from commercial ventures, which are most effectively summarized in reviews from events such as the Consumer Electronics Show (CES) and the Society for Information Display, Display Week.

As proposed in this paper, dual-sided transparent displays, visualized in **Fig. 1**, provide an augmented surface enabling information sharing to two viewers simultaneously while maintaining a transparent window between each viewer. To produce such a device, there are three immediate challenges, illustrated in Fig. 1d-f, that we describe as the 3-O's, occlusion, obstruction, and obversion. The first O is occlusion, which describes how one object hides another behind itself.[26,27] With regards to the display, the device cannot allow information shown on one side to obscure information that is shown on the other. Therefore, stacking two layers of transparent OLED panels would not suffice to create a dual-sided display. The second O is obstruction. This implies that the device should operate as a window when not in use, and an interlayer material between the displays would block the view and is not desired. Finally, the third O is for obversion, which stands in place for the description of bleed-through as described earlier. Obversion of an image, presented on an initial side, should be prevented from being shown on the opposite of the display in the form of a mirrored image, thus providing unintelligible information to an unintended viewer. If the challenges proposed by these three O-'s can be overcome, an effective augmented reality device can be assimilated into any window based viewing system.

Computer science and human-machine interface research groups have investigated human interactions and experiences utilizing dual-sided displays by projecting information onto half mirrors or tinted windows at specific angles to minimize light transmission.[25,28] Another group utilized liquid crystal display and polarizing components to achieve their goals.[25] These existing demonstrations provide unique solutions utilizing off-the-shelf components to create their displays, thereby limiting the fundamental advancement of the device specifications. To create a high-performance commercial device, the base technology should be modified to achieve a high brightness, high transparency, high contrast ratio device. As a launching point, for the work presented herein we recognize the efforts of a "single" sided, transparent display created from an edgelit light plate, where the liquid crystal panel acts as the conductor for the light and ITO electrodes are utilized to activate the pixels from their off to on state.[15,28–32]

In this paper, we demonstrate an early prototype of a dual-sided transparent display utilizing liquid crystal technology through a light plate. The device demonstrates imagery on both sides of the display without any crosstalk and does not suffer from occlusion, obstruction, or obversion. Moreover, the device retains a transparency of 65% and an average brightness of 16 cd/m$^2$.

## 2. DESIGN

Three research stages of display cells were utilized in the development of the dual-sided transparent display. Each stage is slightly different but maintains three functional components, an LED light source that shuttles light through the light plate display, the liquid crystal that scatters the light, and the electrodes that control the liquid crystal modulation. The three stages of the display shown in **Fig. 2** were implemented in order to reduce complexity and cost throughout the material research design phase.

The Stage 1 display, shown in Fig. 2a, is the easiest and cheapest display to fabricate, thereby allowing early-stage parametric testing of the polymer stabilized liquid crystal matrix. The Stage 1 device is a single-pixel, waveguide display that portrays identical images on either side of the display. It should be noted that the Stage 1 display is not the proposed dual-sided display as it creates mirrored images on both sides of the display, rather than two unique non-interfering images. This prototype device consists of indium tin oxide (ITO) patterned on two glass substrates, separated by micron sized spacers and filled with a liquid crystal matrix. The ITO is patterned into electrodes that are 25 μm in width (active area) and a non-ITO strip of 250 μm (inactive/transparent area); a schematic of the device can be found in Supplementary **Fig. S1**. Next, an alignment layer of polyamide (PI) SE2171 from Nissian Chemical is coated on the ITO and then rubbed to create a homogeneous alignment. The cell is created by joining both patterned pieces of glass and creating a cell gap that is uniformly controlled at 2 μm using spherical spacers. The cell gap of 2 μm was chosen after evaluation of the cell performance when utilizing 2, 3, and 4 μm spacers see Supplementary **Fig. S2**. The cells were filled with a mixture of a monomer, the nematic liquid crystal BOE-5, and a photoinitiator Benzoin methyl ether (BME). After the mixture is injected into the display cell, the cell is irradiated by ultraviolet (UV) light, and the monomers are polymerized to form a polymer network within liquid crystal matrix.

In addition to the Stage 1 single pixel device, a Stage 2 and Stage 3 device are formed that enable unique non-interfering images to be portrayed on either side of the display. The stage 2 prototype, as shown in Fig. 2b, is a 2" x 2" display that consists of 16 individually controllable pixels in a 4x4 matrix. The Stage 3 display is a 4.5" x 4.5" display that consists of 100 individually controllable pixels in a 10x10 matrix, Fig. 2c, with the same resolution per pixel as the stage 2 display. Each single pixel contains 31 x 31-pixels that aim out the front of the display and another set of 961 pixels that aims out the rear of the display; refer to Fig. 2d. While Stage 2 and Stage 3 prototype devices have resolutions of only 16 and 100 individually controllable pixels, respectively, these designs were carefully selected to balance cost when prototyping the devices; in the future, each sub-pixel (shown in Fig. 2d) can be independently controlled utilizing a specialized electronic control unit. To protect from bleed-through and therefore the obversion obstacle, both Stage 2 and Stage 3 displays involve a more complicated panel, layer stack up.

A cross-section of both Stage 2 and Stage 3 devices is shown in Fig. 2e. The displays operate by injecting light from an LED array on the left side of the diagram into the waveguide. Within the waveguide, a polymer-stabilized liquid crystal is present, which either allows the light to pass through in the off-state or scatters light perpendicularly out of the display in the on-state due to the oscillating reorientation of the liquid crystal. The function of the polymer network is to create a polydomain liquid crystal structure which enables this scattering when the voltage is applied and utilizes elasticity to quickly relax the device to its off-state. An interleaved matrix of ITO electrodes is arrayed, as shown in the top-down view in Fig. 2f, to control the on/off-state of individual pixels in the display matrix when a voltage bias is applied. Common ground ITO electrodes are distributed on the top side of the display, and two sets of activation electrodes, one for creating an image on the front of the display, and the other for the back, are interdigitated on the bottom side of the panel. An individual unit cell is shown in a perspective-view in Fig. 2g and is denoted as a dashed line in Figs. 2e and 2f. Since this is a multipixel active-matrix display, each of the unit cells is divided into three parts: one active region for the inside viewer, one active region for the outside viewer, and an inactive region to act as a transparency window. In each unit cell, a black mask is applied to one sub-pixel on one side of the waveguide to block light from escaping; this mask is key to providing light to only the desired side and preventing light from leaking to an unintended viewer. The size of the black mask is designed to be large enough to block all the light from a dedicated subpixel but small enough to maintain sufficient transparency of the entire panel. Molybdenum/Aluminum/Molybdenum (Mo/Al/Mo) wiring is utilized to connect to terminals at the edges of the ITO strips, where Mo serves as an adhesive layer.[33] The same methods that are utilized for the liquid crystal, that is applied to Stage 1 devices, is also used in filling the Stage 2 and 3 systems. The cell thickness is 3 μm and is controlled by styrene posts. Similarly, the operation of the device is achieved by injecting light via an LED into the sides of the display, and a 1kHz voltage is applied across the electrodes, which rotates the liquid crystals and induces scattering in the display.

## 3. RESULTS

In the development of the display, we structured parametric studies of the Stage 1 devices, Fig. 2a, to investigate key controlling factors of the display, while utilizing the low cost research prototype. The results from these parametric studies are provided as plots of applied-voltage versus light intensity scattered from the display in Figs. 3a-c, while Fig. 3d-f provide bar chart analyses of these studies in the form of contrast ratio and saturation voltage. The contrast ratio (CR) is a crucial parameter of image quality for a liquid crystal display, especially a waveguide display, which is defined as the ratio between on- and off-state light intensities, $CR = I_{max}/I_{min}$. Therefore, low light leakage at the off-state and high light intensity at the on-state is desirable for high contrast and enhanced brightness of a display. The second parameter, the saturation voltage $V_{sat}$, is defined as $V_{sat} = I_{min} + 0.9(I_{max} - I_{min})$ and in this paper it refers to the voltage level required to fully activate (or saturate) the liquid crystals in a pixel, where $I_{max}$ is the maximum intensity of light scattering from the voltage-intensity curve. In addition to the CR, the saturation voltage is a key determinant in selecting which configuration moves to the next device stage. In our case, we prefer a lower voltage for the saturation voltage as this implies a lower operating power.

As an initial study, we investigate the influence of the monomers on the electro-optical properties of the display in the Stage 1 device. The applied voltage-dependent light intensity curve of the displays with various monomers combinations are shown in Figs. 3a,b. Upon analysis of the graph in Fig. 3a, it is evident that the Stage 1 device fabricated with monomer HCM-009 exhibits notable light scattering capabilities with a maximum light intensity ($I_{max}$) slightly exceeding 5 a.u., accompanied by an off-state intensity ($I_{min}$) of approximately 0.22 a.u.

Furthermore, this configuration exhibits maximum brightness when the display is in the on-state and reduced light leakage from the display in the off-state, resulting in a contrast ratio of approximately 24, thus establishing HCM-009's superiority over the other monomers investigated herein. The saturation voltage associated with HCM-009 measures 10.5 V. The next best monomer is RM-257. The electro-optical performance of monomer RM-257 is moderate, with a maximum light intensity, $I_{max}$= 4.1 a.u., an off-state intensity of $I_{min}$=0.26 a.u., a CR=15.9, and $V_{sat}$=12 V. Though the saturation voltages of the displays prepared using monomer BAB-6 and RM-82 are low, ~9 V, other electro-optical properties such as the on/off-state light intensity and contrast ratio are less satisfactory. Based on these results, we selected HCM-009 for further prototypes.

After studying the electro-optical properties of the different monomers, we optimized the concentration of the best-performing monomer, HCM-009. The applied voltage versus light intensity curve at various HCM-009 concentrations is shown in Fig. 3c and Fig. 3d provides the corresponding CR and saturation voltages. As the monomer concentration increases, the saturation voltage increases until it reaches a maximum at 7%. Meanwhile the best CR is achieved at a concentration of 6%, thereby maximizing the light intensity relative to the off-state.

With the optimized 6% concentration of HCM-009 selected, a parametric study of the UV light intensity needed for polymerization was performed; the results depicted in Figs. 3e,f, showcase the relationship between light intensity and applied voltage. Notably, both the off-state light intensity and the maximum light intensity show an increase in UV light intensity up to 8 mW/cm², but then subsequently declined. Thus, the UV polymerization intensity of 8 mW/cm² is optimal. Given these parameters, a contrast ratio of 24 is achieved. Furthermore, it is worth noting that the driving voltage initially decreased from 10.4 V to 9.9 V at an intensity of 6 mW/cm² before undergoing a subsequent increment.

To conclude the studies regarding the Stage 1 device, the switching time or switching speed was measured for a variety of configurations of the devices. In general, the switching speed did not show remarkable improvements from the switching time shown in Fig. S3, which was measured for the 6% HCM-009 device, with a UV curing intensity of 8 mW/cm². The cell turn-on time is 1 ms, and turn-off time is ~ 2 ms, which allows for time sequential operation of the device to generate a full-color image. Full time sequential operation allows the display to operate without having to rely on color filters embedded into the device, which would significantly decrease the transparency of the display.

After the monomer selection and process optimization in the test cells was complete, the Stage 2 and Stage 3 prototype displays shown in Fig. 2 were fabricated. In Fig. 4a, an image of the Stage 2 prototype against a blue sky at the Kent State campus is demonstrated to show the transparency of the display. In Fig. 4b, a comparison between the transmission spectrum in the visible regime is shown for the display and a single pane of glass for reference. The transparency of the Stage 2 device relative to air is approximately 65% over the visible range, indicating its optical clarity when the device is inactive. The loss of light relative to the pane of glass can be mainly attributed to the presence of the black mask and the ITO electrodes. The total loss of transparency due to the black mask can be calculated as 18%. The voltage dependent electro-optical properties were also measured for the display. As shown in Fig. 4c, measurements of the scattered light intensity are obtained from the middle of the device, between the two input light sources on either side of the LCD panel. The plots demonstrates a noticeable difference between the front and rear sides of the display in terms of light intensity, with the front side exhibiting slightly higher overall light intensity. Based on these measurements, we determined the contrast ratio of the display to be approximately 4.34; as mentioned previously this value indicates the distinction between the brightest and darkest areas of the display. Additionally, we measured the brightness of the display to be approximately 16 cd/m², signifying its luminous intensity. These findings provide valuable insights into the optical performance of the Stage 2 display and its suitability for further evaluation and optimization.

Finally, a Stage 3 display cell was demonstrated and powered by a driver board to display different images utilizing time-sequential processing. The display can be seen in various conditions as shown in **Fig. 5** and Supplementary **Video S4**. Fig. 5a provides an image of the display with the LED array integrated into a housing. The image is taken indoors under ambient light conditions. The housing supports the flex ribbon and associated board that are attached to the display, and also serves to secure the LEDs tight to the side of the waveguide display. Figs. 5b and 5c show the front (Side A) and back (Side B) views of the display, respectively. The display images shown on either side do not interfere with each other, therefore, in combination, the images confirm the operation of a true dual-sided transparent display system. Moreover, the images in the dark regions are, in fact, transparent as the images were taken in a dark room. The transparency of these regions can be visualized in the Supplementary Video S4.

## 4. CONCLUSIONS & DISCUSSION

In this article, we demonstrated a dual-sided transparent display. The display operates as a light plate with LEDs injecting light into the side of the display. By parametrically studying the monomer type, monomer concentration, and UV intensity, the device was optimized to achieve the highest contrast ratio of ~24 and the lowest driving voltage of about 10.5 V for the single-pixel display. At present, the multi-pixel display demonstrates a transparency of 65%+ over the visible window, a contrast ratio of approximately 4.34, and a brightness level of 16 cd/m². The relative loss in contrast ratio from the Stage 1 to Stage 3 prototype can be attributed to the presence of the black mask and the pixel arrangement of the device.

A variety of enhancements are possible as a next step. In the current state of the device, the full potential of the resolution of the device is underutilized. Each pixel within the 10 x 10 pixel device is composed of many smaller sub-pixels that are each 90 μm x 90 μm in size. These smaller sub-pixels could be individually controlled, however, the cost of developing circuitry/software for these early prototypes would be extremely high. Connecting each sub-pixel could provide an opportunity to test the brightness and contrast ratio of a high-resolution device while keeping upfront investments low. Individually controlling these sub-pixels is estimated to increase the display resolution by more than 300 times. Further optimization of the pixel size and coupling of the LED light into the light plate waveguide is expected to further increase brightness and improve the contrast ratio of the device. Specifically, in the current setup, it was necessary to be able to easily switch the LED from one device to another in modular fashion for fast prototype evaluation. However, as a result, the total coupling of light into the device from the LED was deemed inefficient based on observed light leakage from the edge of the display during prototype tests. In the future, additional optimization of the coupling of the light into the device is expected to further increase the brightness of the display significantly. We envision this dual-sided transparent display technology as having applications in a variety of environments including informative displays such as office and vehicular windows, enhanced customer-facing interactions such as ridesharing interfaces, checkout lines at ice cream, fast food, or translation services, and user-user augmented discussions such as in-office communications, and gaming platforms: think battleship-like games.

**Acknowledgements**

This study was supported by funding from MIRISE Technologies and Toyota Motor Corporation.

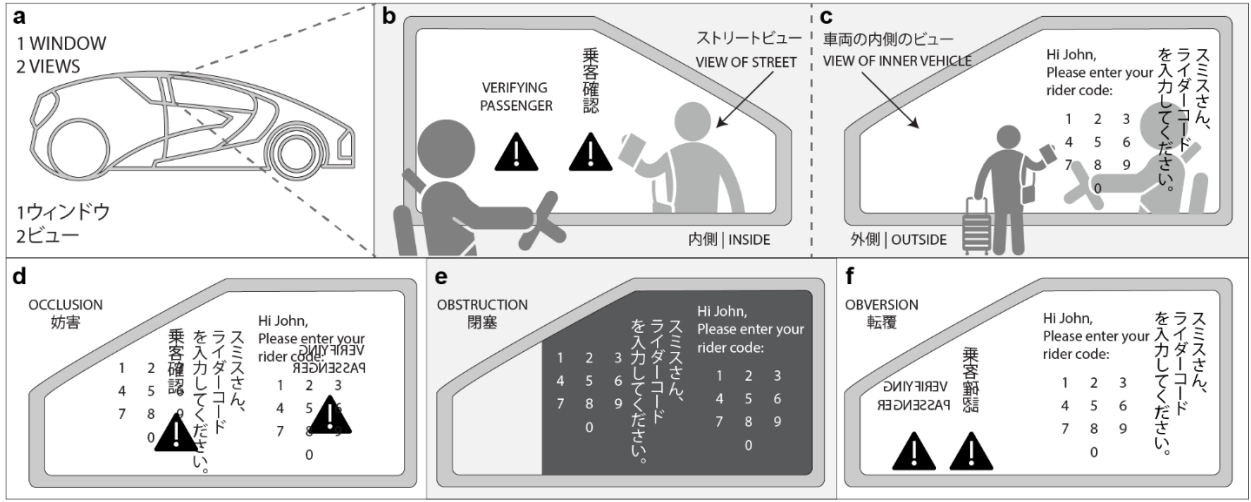

**Fig. 1: Concept for dual-sided transparent display**.

**a** A vehicle installed with a dual-sided display could utilize one window to provide two unique augmented scenes. **b** A depiction of what the driver would see from the inside the vehicle on the interior side of the window, are shown on the interior of the window, describing what the driver would see from inside the vehicle while the majority of the screen remains transparent allowing you to see the passenger. **c** A depiction of what the passenger would see approaching the vehicle. Note although the same window is used, graphics on either side occlude the image, as shown in **d**. Most technologies **e** Although an interlayer could be used between two screens, this type of occlusion is not possible for vehicular applications. **f** It is possible to use software to defeat occlusion using a single screen, however, the obverted imagery would not be desired by a customer.

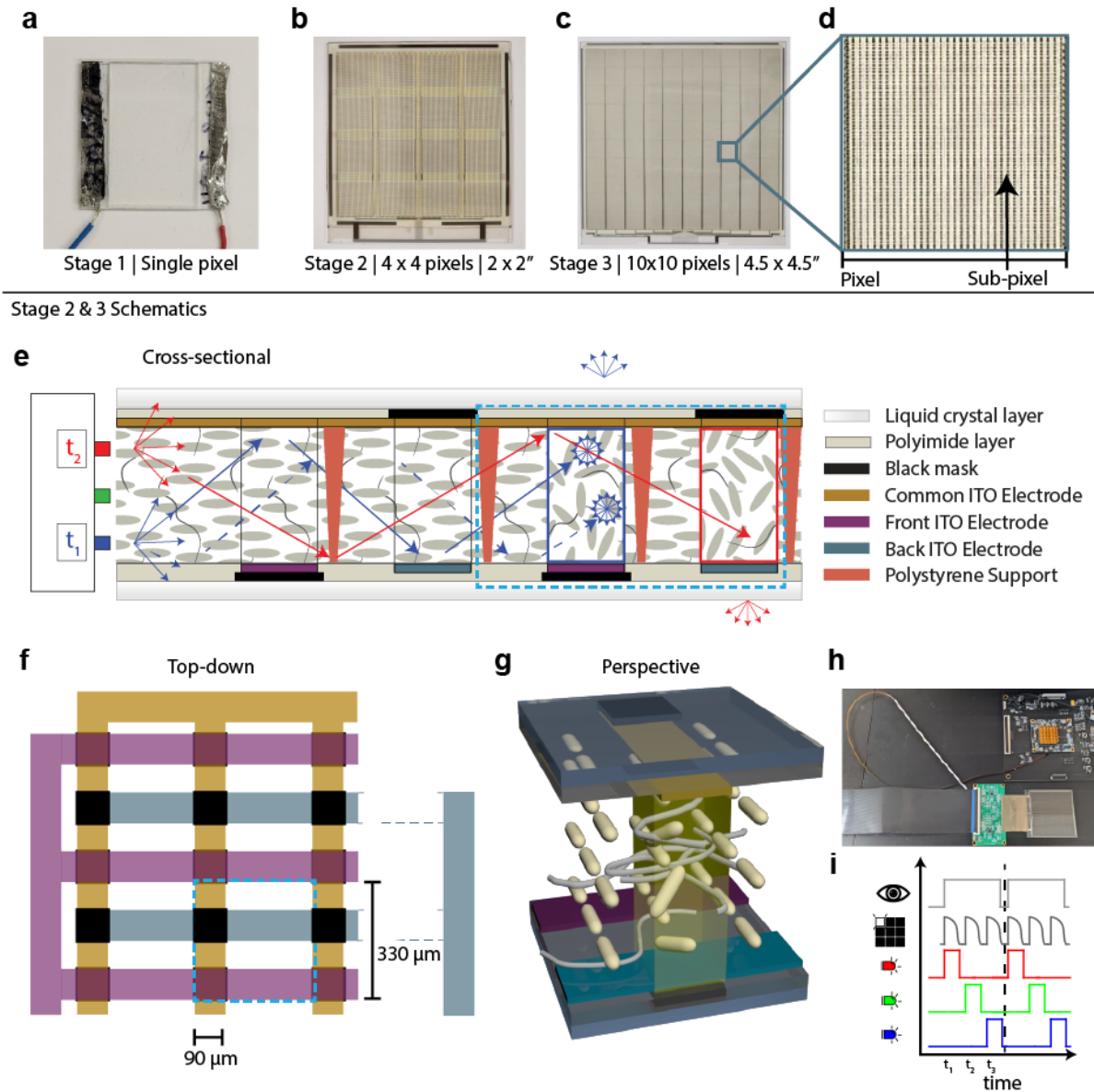

**Fig. 2: Design and operation of the dual-sided transparent display**.
**a** The Stage 1 prototype is a single pixel device that allowed for early stage testing of liquid crystal cell testing. **b** The Stage 2 prototype has 16 total pixels that can be controlled independently. **c** The Stage 3 device is composed of 100 individual pixels. Control wiring connects to the lower portion of each device. **d** Zoom in of a single pixel. Each pixel is composed of a set of 33 x 33 unit cells. **e** A cross sectional schematic of the dual-sided display, with 2 sample time frames for the time-sequential processing. The dashed line represents a single unit cell that is composed of 2 sub-pixels, one that faces the driver and one that faces the passenger as schematically described in Fig 1. Scattering is blocked from the undesired side by a black mask. **f** A top-down schematic of the electrodes for Stage 2 and Stage 3 devices. Where the labels for the electrodes are shown in **e**. **g** A perspective schematic that depicts the liquid crystal and polymer network. Here a single sub-pixel, highlighted in yellow, is scattered within a unit cell. **h** A top-down image of the dual-sided display, its driver, and its LED. **i** A pictorial description of describing time-sequential color graphics.

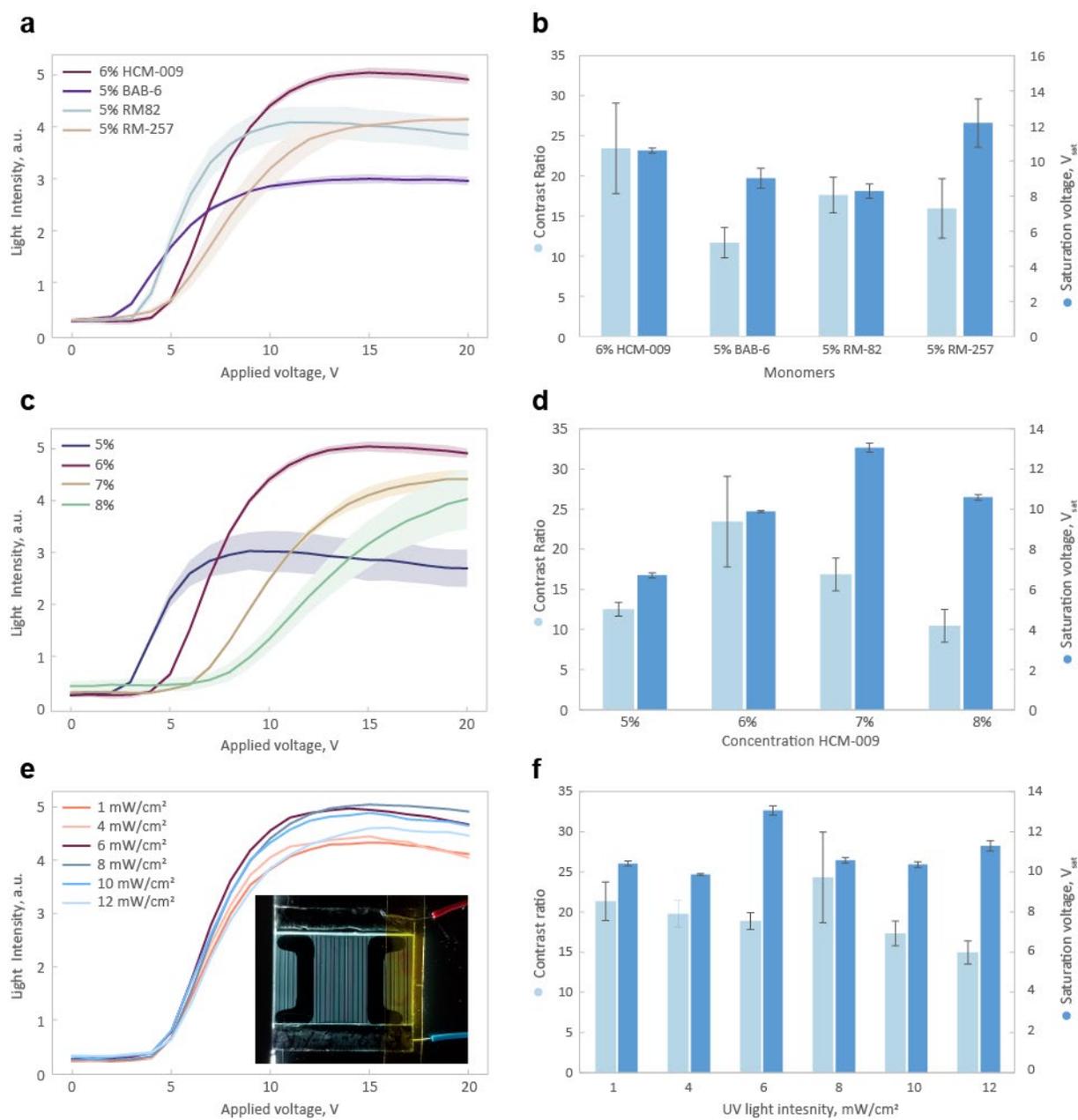

**Fig. 3: Optimization of cell conditions in Stage 1 devices.**

Studies of intensity vs voltage were applied for **a** Various monomer concentrations tested; **c** Varied concentrations of HCM-009; **e** Varied UV polymerization intensities using HCM-009. **b,d,f** Contrast ratio and saturation voltage for the corresponding graphs shown in **a**, **b**, **c**.

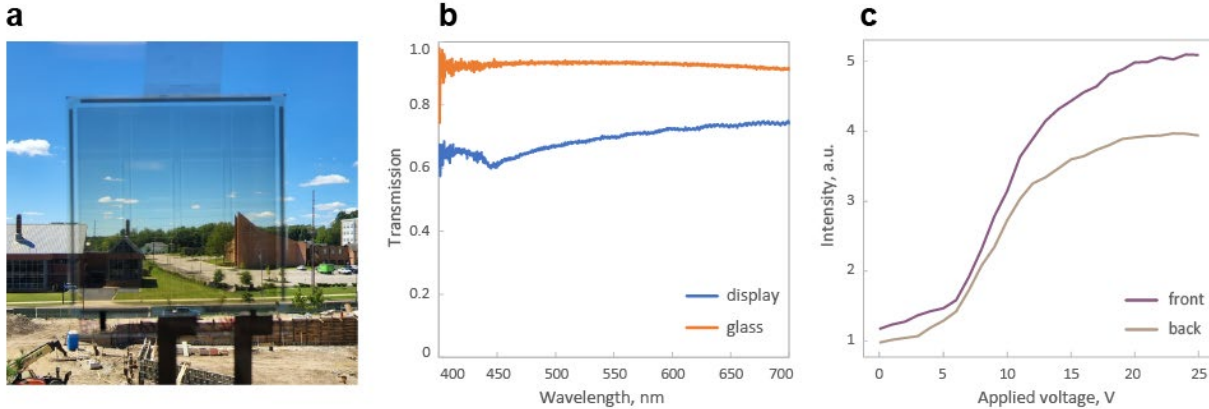

**Fig. 4: Prototype characteristics of the dual-sided, liquid crystal display.**

**a** Image of the Stage 2 display demonstrating the transparency of the device. The wiring for the device would connect at the bottom and the LEDs would illuminate from the left and ride of the display. **b** Transmission spectra of the display is shown relative to a single glass pane. The main factors of transparency loss in the display are due to the black mask, the ITO, and the first air to glass reflection. **c** Plot of intensity vs voltage for the 4 x 4 pixel display.

**a**

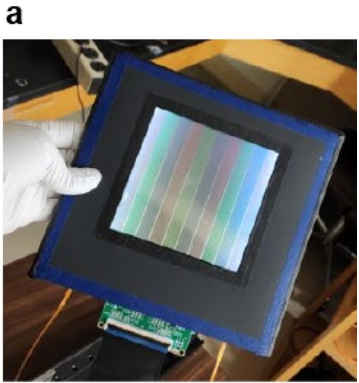

**b**

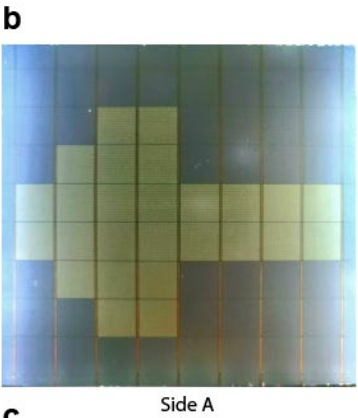

Side A

**c**

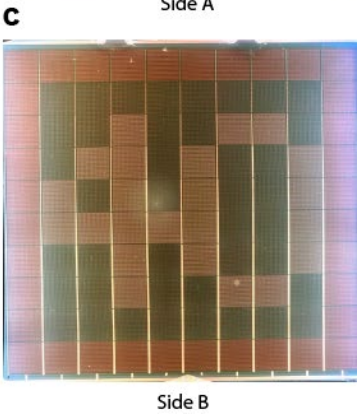

Side B

**Fig. 5: Prototype characteristics of the dual-sided, liquid crystal display.**

**a** Image of the display in a housing while operating in ambient light conditions. The housing holds two sets of LEDs illuminating the waveguide display (not shown). **b** The front-side of the display shows an arrow pointing to the left. **c** Simultaneously the image on the back-side of the display shows a street sign with the number 40 on it. The dark regions on the display correspond to transparent regions, however in order to capture the maximum brightness the images were taken in the dark.

**Supplementary Information**

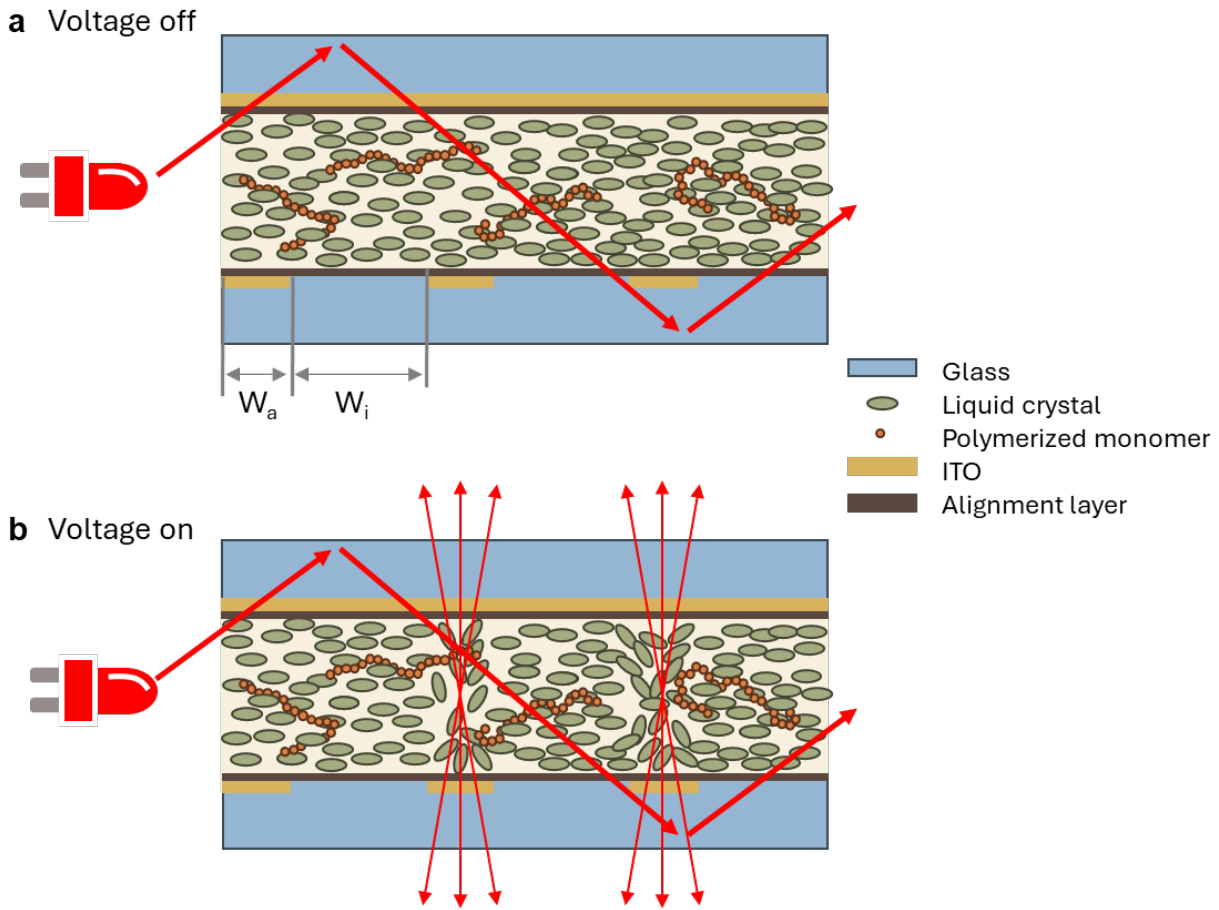

**a** Voltage off

**b** Voltage on

Glass
Liquid crystal
Polymerized monomer
ITO
Alignment layer

**Figure S1: Operation of the Stage 1 prototype devices.** These devices were utilized to study the effects of different types of monomer, their concentration, and UV intensity. The device is comprised of two glass slides that form the cell. One glass slide has a layer of ITO that is completely un-patterned. The other side of glass has a patterned layer of ITO. The ITO pattern is a series of stripes that have widths that are 25 μm ($W_a$), and inactive regions that are 250 μm ($W_i$), for a total period of 275 μm. The ITO thickness is ~150 nm. Above the ITO, is an alignment layer on both the patterned and un-patterned sides of glass. The alignment layer is a layer of polyimide that is rubbed perpendicular to the direction of the input light. The liquid crystal is then filled into the cells.

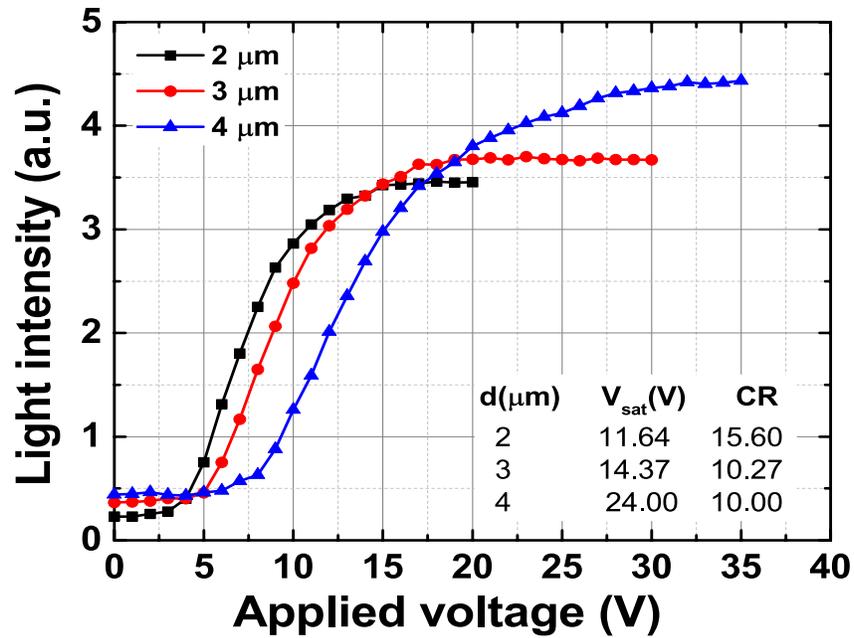

**Figure S2: Studies of intensity vs voltage were applied for various cell thicknesses (inset: CR & Vsat)**

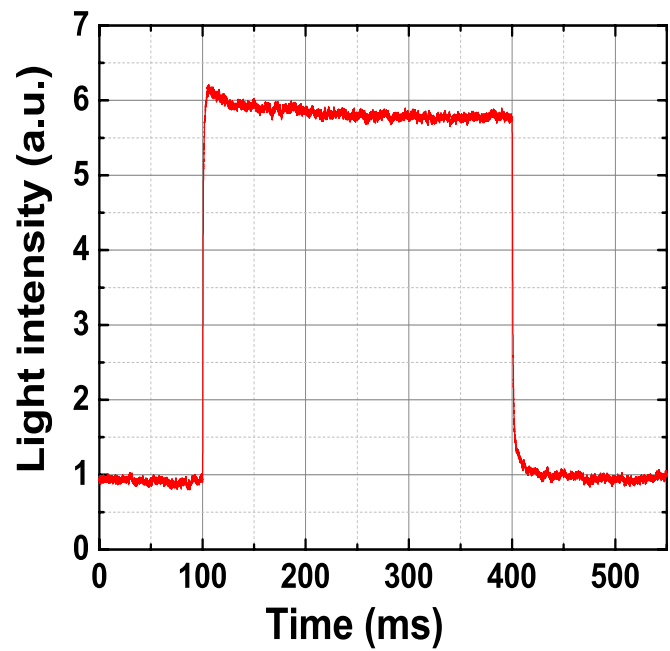

**Figure S3: Light intensity vs. time curve of the display prepared by 6% HCM-009**